\documentstyle[aps,twocolumn,psfig]{revtex}
\begin{document}
\draft
\wideabs{
\title{Adsorbed states  of a long - flexible polymer chain}
\author{ Yashwant Singh, Debaprasad Giri$^\dagger$ and Sanjay Kumar}
\address{ Department of Physics, Banaras Hindu University, Varanasi 
221 005, India, \\ 
$^\dagger$Present address: Centre for Theoretical Studies, IIT, 
Kharagpur 721 302, India.} 
\date{\today}
\maketitle
\begin{abstract}
A phase diagram for a surface-interacting long flexible
polymer chain in a two-dimensional poor solvent where 
the possibility of collapse exists is determined using
exact enumeration method. A model of a self-attracting 
self avoiding walk (SASAW) on a square lattice was considered 
and up to 28 steps in series were evaluated. A new adsorbed 
state having the conformation of a surface attached globule 
is found. Four phases (i) desorbed expanded, (ii) desorbed 
collapsed, (iii) adsorbed expanded and (iv) surface attached
globule are found to meet at a point on the adsorption line. 
\end{abstract}                                                               
}
\narrowtext 

The behaviour of a long flexible polymer chain near an 
impenetrable surface is a subject of considerable experimental
and theoretical importance \cite{1,2,3}. This is because such a 
chain may exhibit a phase diagram characterized by many different 
universality domains of critical behaviour. The subtle competition 
between the gain of internal energy and corresponding loss of entropy 
at the surface may lead to the possibility of the coexistence 
of different regimes and multicritical behaviour. An attracting 
surface may lead adsorption - desorption transition from the 
state when the chain is mostly attached to the surface, to the 
state of detachment when the temperature is increased. This behaviour
finds applications in lubrication, adhesion, surface protection etc.

The essential physics associated with the behaviour of a surface 
interacting polymer chain in a good solvent where monomer-solvent 
attraction is greater than the monomer-monomer attraction, is derived 
from a model of self-avoiding walk (SAW) on a semi-infinite lattice. 
If the surface is attractive, it contributes an energy $\epsilon_a$ 
($< 0$) for each step of the walk along the lattice boundary. This leads 
to an increased probability characterized by the Boltzmann factor 
$\omega = \exp(-\epsilon_a/k_{\beta}T)$ of making a step along the wall, 
since for $\epsilon_a < 0$, $\omega > 1$ for any finite temperature $T$ 
($k_{\beta}$ is the Boltzmann constant). Because of this the polymer
chain becomes adsorbed at low temperatures on the surface while at 
high temperatures all polymer conformations have almost same weight
and non adsorbed (or desorbed) behaviour prevails. The transition 
between these two regimes is marked by a critical adsorption 
temperature $T_a$, with a desorbed phase for $T > T_a$ and adsorbed 
phase for $T < T_a$. At $T = T_a$ one may define the crossover exponent
$\phi$, as $M \sim N^{\phi}$, where $N$ is the total number of steps 
and $M$ the number of steps on the surface. The transition point 
$T_a$ is a tricritical point \cite{3}. Both the surface and the bulk 
critical exponents have been calculated using renormalization group 
methods \cite{4}, exact enumeration methods and Monte-Carlo simulations 
\cite{5,6,7,8}. For a two dimensional system exact values of the exponents 
have been found by using conformal invariance \cite{9}.

The situation, is, however, different when the surface interacting 
polymer chain is in a poor solvent where monomer-monomer attraction 
dominates over the monomer-solvent attraction. As is well known, a 
long-flexible polymer chain in a poor solvent exhibits a transition 
from a compact globule (collapsed state) to a expanded state when the 
temperature is increased. Above the critical $\theta$ temperature (often
referred to as the $\theta$-point) the chain behaves as it  would in a
good solvent and below this temperature it behaves like a compact
globule. At the $\theta$-point the chain behaviour is described by 
a tricritical point of the $O(n)$ ($n \rightarrow 0$) spin system
\cite{10}.  However, when the chain is in the vicinity of an impenetrable 
surface the competition between the monomer-monomer attraction and the 
surface-monomer interaction gives rise to many new features. Attempts 
have been made to study these features using several approaches 
\cite{3,7,8,11}.

For two-dimensions the transfer matrix method has been used for a 
directed polymer  chain \cite{12} whereas for the nondirected (isotropic)
version the exact enumeration method has been used \cite{8}. In both
cases, three phases, desorbed expanded, desorbed collapsed and a single
adsorbed phase have been predicted. However, the true nature of the phase
diagram remained unknown. For three dimensions, the Monte-Carlo simulations 
method \cite{13} has been used for a finite length ($\sim$ 100) chain 
which led to a phase diagram containing  four phases; desorbed expanded (DE), 
desorbed collapsed (DC), adsorbed expanded (AE) and adsorbed collapsed (AC). 
The phase diagram shows a phase boundary between the AE and DC phases 
leading to two points on the phase diagram where three phases coexists(triple
point). However, the phase diagram found by the exact enumeration technique 
\cite{14} has many features which are different from  that found in ref.
\cite{13}. This indicates the possibility of a richer phase 
diagram than has been realized so far. In view of this we reexamine the 
problem of simultaneous adsorption and collapse of a linear polymer chain
on a square lattice and investigate the phase diagram and critical parameters
using the exact enumeration technique. We prefer this technique because 
in this case the scaling corrections are correctly taken into account 
by a suitable extrapolation scheme. As shown by Grassberger and Hegger 
\cite{5}, to achieve the same accuracy with the Monte Carlo method one has 
to consider a polymer chain of about two orders of magnitude longer than in 
the exact enumeration method.

We consider SASAW on a square lattice restricted to half space 
$Z \ge 0$ (impenetrable hard wall). Walk starts from the middle 
of the surface. Let $C_{N,N_s,N_m}$ be the number of SAWs with
$N$ steps, having $N_s$ $(\le N)$ step on the surface and $N_m$
nearest neighbor. We have obtained $C_{N,N_s,N_m}$ for $N \le 28$ 
for square lattice by exact enumeration method.  

Now we consider the interaction energy $\epsilon_a$ associated
with each walk on the surface and $\epsilon_m$ for
monomer-monomer interaction. Partition function of the attached
chain is 
\begin{equation}
Z_N (\omega, u) = \sum_{N_s,N_m} C_{N,N_s,N_m} \omega^{N_s}
u^{N_m} 
\end{equation}
where $\omega = e^{-\epsilon_a/kT}$ and $u = e^{- \epsilon_m/kT}$.
$\omega > 1$ and $u >1$ for attractive force. Reduced free
energy for the chain can be written as 
\begin{equation}
G (\omega, u) = \lim_{N\rightarrow \infty} \frac{1}{N} \log
Z_N(\omega,u) 
\end{equation}
In general it is appropriate to assume that as $N\rightarrow
\infty$ 
\begin{equation}
Z_{N}(\omega,u) \sim N^{\gamma - 1} \mu(\omega,u)^{N} 
\end{equation}
where $\mu(\omega,u)$ is the effective coordination number and
$\gamma$ is the universal configurational exponents for walks
with one end attached to the surface. The value of
$\mu(\omega,u)$ can be estimated using ratio method \cite{15}
with associated Neville table. From equations (2) and (3) we can
write 
\begin{equation}
\log \mu(\omega, u) = \lim_{N\rightarrow \infty} \frac{1}{N} \log
Z_N(\omega,u) = G (\omega, u)
\end{equation}
\par $Z_N(\omega,u)$ is calculated from the data of $C_{N,N_s,N_m}$
using equation (1) for a given $\omega$ and $u$. From this we
construct linear and quadratic extrapolants of the ratio of
$Z_N(\omega,u)$ for the adjacent values of $N$ as well as the
alternate one. Results for alternate $N$ give better
convergence. When $u = 1$ and $\omega = 1$ the value of $\mu$ is
found to be 2.638 which is in very good agreement with the value
given in ref. \cite{2,7}.

The value of $\omega_c (u)$ at which polymer gets adsorbed for a
given value of $u$ is found from the $(i)$ plot of $G(\omega,u)$ which remains
fairly constants until $\omega = \omega_c$ and increases 
 consistently as a function of $\omega$, for $\omega \ge \omega_c $ $(ii)$
from the plot of $\partial^2 G(\omega,u) / \partial {\epsilon_a}^2$ at constant
$u$ and $(iii)$ from the plot of $\gamma^0 - \gamma_1$ ( see Eq.(6) and 
discussions which follow it) as a function of $\omega$ for
 different $N$. The value of
$\omega_c$ found from the plot of $G(\omega,u)$ is slightly lower than the
peak value of $\partial^2 G/ \partial {\epsilon_a}^2$. It is, however, observed
that as $N$ is increased from $22$ to $28$  the peak value shifts to smaller
$\omega$ and appears to converge on the value of $\omega$ found from $G(\omega,u)$ plot.
We therefore choose the value of $\omega_c$ found from the plot of $G(\omega,u)
$ and determine lines $\omega_c(u)$ and $\omega_{c1} (u)$ (see Fig.1)
 by this method.
For $u = 1$, the value of $\omega_c$ is 2.050  which is in very
good agreement with the value (= $2.044\pm0.002$) reported in ref.
\cite{8}.  Similarly the phase boundary separating the extended and  
collapsed phases is calculated from the plot of $G(\omega,u)$ as a 
function of $u$ for a given $\omega$. However, transition point $u_c$ is 
located more accurately from the peak of $\partial^2 G(\omega,u) / \partial {\epsilon_m}^2$
at constant $\omega$. For $\omega = 1$, the value of $u_c$ is
1.93 which is in good agreement with the value found 
by Foster et al \cite{8} and the Monte Carlo results 
(= 1.94 $\pm$ 0.004) \cite{5}. 
The method is found to work for all values of 
$\omega$ {\it i.e} in both the bulk and the adsorbed regimes.
However, as $\omega$ is increased the values of $G(\omega,u)$ do not 
remain as smooth as at lower values of $\omega$, therefore introducing 
some inaccuracy in the value of $u_c$. The estimate of this inaccuracy 
is of the order of $5\%$  for $\omega > 4$. We therefore conclude that the
$u_c$ and $\omega_{c2}$ (Fig.1) lines are determined with reasonable accuracy.

The surface critical exponent $\gamma_1^{N,k}$ can be calculated
using the relation:
\begin{equation}
\gamma_1^{N,k} = \frac{\log(Z_N/Z_{N-2}) - 
k\log(\mu)}{\log(N/(N-k)} + 1
\end{equation}
where subscript ``1" indicates the corresponding  quantity of 
the surface (with one end of the polymer is attached to the surface).  
If we assume that $\mu$ does not depend on $\omega$ and polymer 
is in desorbed phase, then we can calculate the quantity
$\gamma^0 - \gamma_1$ from the above equation
\begin{equation}
\gamma^0 - \gamma_1 = \frac{\log(Z_N^0 Z_{N-2} / Z_{N-2}^0
Z_N)}{\log(N/N-2)}
\end{equation}
where superscript ``0" indicates the corresponding quantity of
the bulk ({\it i.e} without surface).  In this case one
calculates $\gamma^0 - \gamma_1$ for different $N$ using above
equation and plot it as a function of $\omega$. The location of
adsorption point $\omega_c$ can be determined from the intersection
of successive approximation to $\gamma^0 - \gamma_1$ in the limit
$N \rightarrow \infty$. This method, however, fails beyond 
the $\theta$-point and reproduces closely the adsorption phase
boundary below the $\theta$-point found by the method discussed
above.

The phase diagram shown in Fig. 1 has four phase boundaries instead
of three as reported in earlier work \cite{8}. The $u_c$ line separates
the expanded and collapsed phases. This line remains straight 
and parallel to $\omega$-axis in the bulk. This result is in agreement 
with that of ref. \cite{8}. The special adsorption line $\omega_c$ 
separates adsorbed expanded (AE) phase from that of  desorbed 
expanded (DE). Beyond $\theta$-point we have two boundaries $\omega_{c1}$ 
and $\omega_{c2}$. The line $\omega_{c1}$ separates the desorbed collapsed
(DC) bulk phase from that of an adsorbed globule (AG) state, whereas
the boundary $\omega_{c2}$ separates the AG phase from the AE phase. 
The point where $u_c$ line meets the special adsorption line $\omega_c$, 
all the four phases AE, DE, DC and AG coexist. The AG phase which exists 
between the boundaries $\omega_{c1}$ and $\omega_{c2}$ for $u > u_c$ 
is essentially a two-dimensional globule sticking to the surface in the
same way as a liquid drop may lie on a  surface. The existance of such
a phase, to the best of our knowledge, is shown for the first time.

\psfig{figure=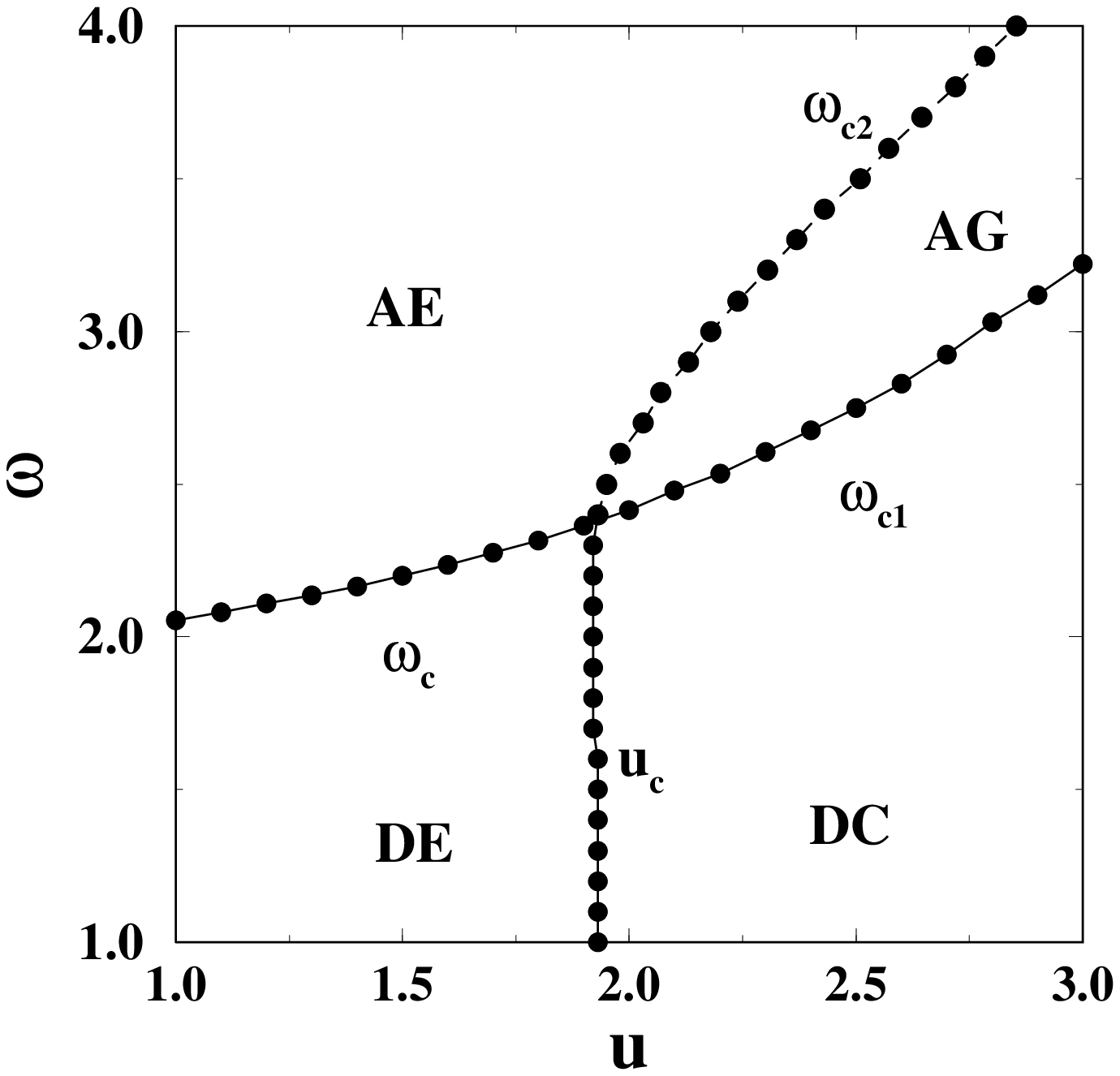,height=4in,width=4in}
\vspace {-1.5in}
{\bf FIG.1: \small The phase diagram of a surface interacting
linear polymer in 2-D space.  $\omega$ and $u$ axes represent,
respectively the Boltzmann factor of surface interaction and
monomer-monomer attraction. Regions marked by AE, AG, DE and DC
represent, respectively, the adsorbed polymer in expanded (
swollen) state and globular state, desorbed polymer in expanded
and collapsed state.}
\vspace {.2in}
                
To find the monomer density as a function of distance normal to the
surface for different regimes of the phase diagram we first found
the terms which make most significant contribution in Eq.(1) for the
walks of 26 steps. For these walks we evaluated the monomers (number
of visited sites) on different lattice layers. The results are shown in
Fig. 2 where we plot the fraction of monomers lying on different layers 
for many values of $u$ and $\omega$ corresponding to different parts 
of the phase diagram. Since the walks always start from the surface, 
there is at least one monomer on the surface for all the cases. In 
Fig. 2(a) $u$ is taken equal to 1.5 which is less than $u_c (1.93)$ and 
therefore it corresponds to expanded state. The values of $\omega = 1.0$ 
and 3.5 correspond, respectively to DE, and AE phases, while $\omega = 2.20$ 
lies on the special adsorption line $w_c$ for $u = 1.5$.  Fig. 2(b) 
shows the change in monomer density as $\omega$ is increased for 
$u=2.5$. In this case $\omega = 1.0$, 3.4 and 3.59 correspond to
DC, AG and AE phases, respectively. For $u = 2.5$ $\omega_{c1} = 
2.75$ and $\omega_{c2} = 3.50$, thus the value $\omega = 3.59$ 
is just above the $\omega_{c2}$ line. The large change in monomer
density distribution when $\omega$ value is changed from 3.4 (slightly 
below $\omega_{c2}$ line) to 3.59 (just above $\omega_{c2}$ line) is
evident and confirms the existence of new (AG) phase.

In Fig. 3 we plot the quantities $<n_s>$ and $<n_p>$ giving the 
average fraction of monomers on the surface and  number of pairs 
and defined as 
\begin{displaymath}
<n_s> = \lim_{N\rightarrow\infty} \frac{\partial\ln G}{\partial\omega}\mid_{u}, 
\;\; \; 
<n_p> = \lim_{N\rightarrow\infty} \frac{\partial\ln G}{\partial u}
\mid_{\omega} 
\end{displaymath}

\vspace {-.5in}

\psfig{figure=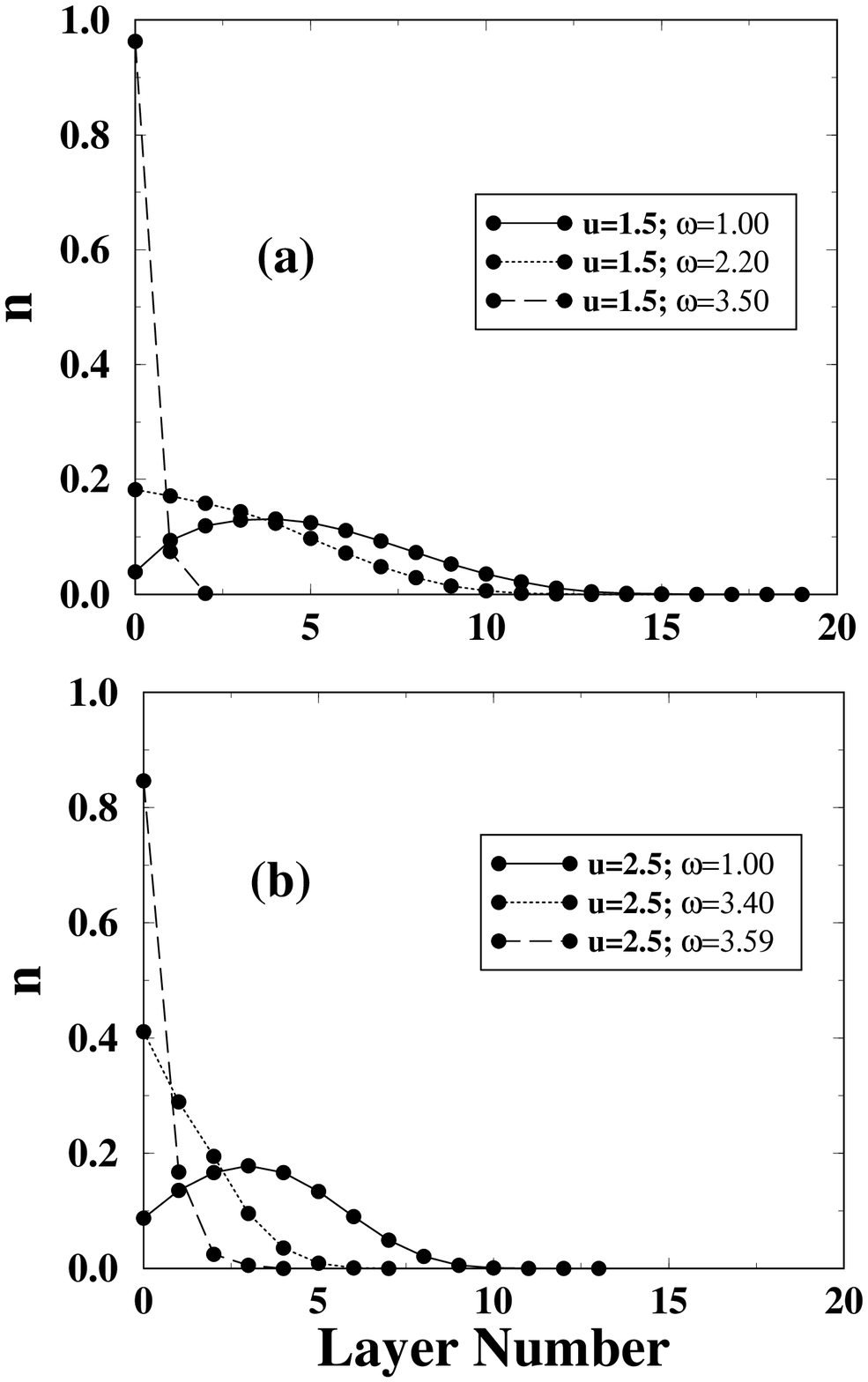,height=5in,width=4in}

\vspace {-.5in}
{\bf FIG.2: \small Fraction of monomers ({\bf n}) on different
layer.}

\vspace {.2in}

The transition points on each curve is marked by dot. While the dot 
on a curve of $<n_s>$ (in Fig. 3(a)) indicate the adsorption 
transition ({\it i.e} point on $\omega_c$ and $\omega_{c1}$ lines
depending on the values of $u$), the dot on a curve $<n_p>$ 
(in Fig. 3(b)) indicates the transition to collapsed state ({\it i.e} 
point on line $u_c$ and $\omega_{c2}$ depending on the values of
$\omega$). We may note that the value of $<n_p>$ for AG phase
is comparable to that in the DC phase. We have found that along the 
line $\omega_c$ and $\omega_{c1}$, $<n_s> = 0.07 \pm 0.004$ and along
the line $u_c$ and $\omega_{c2}$, $<n_p> = 0.5 \pm 0.003$. This 
shows that the error involved in determining the phase boundaries 
is very small.

It is obvious from these results that when the chain gets adsorbed from the 
expanded bulk state ({\it i.e} for $u < u_c$), it acquires a conformation
at $\omega \simeq \omega_c (u)$ such that a small fraction ($\sim 10 \%$)
of monomers get attached to the surface (see Fig. 3(a)) and others are
still in the bulk. Though the chain has formed a layer parallel to 
the surface there are considerable fluctuations in direction normal to
the surface layer. As $\omega$ is increased for the same value of $u$, the
fluctuations along the normal to the surface get suppressed and at 
large $\omega (>> \omega_c)$ the chain lies on the surface with very 
little fluctuations (see Fig. 2(a)). On the other hand, when the adsorbing
chain was in collapsed bulk state, then at $\omega = \omega_{c1}$ the
collapsed chain gets attached to the surface. Here again the number of 
monomers getting attached to the surface are about $10\%$ (see Fig. 3(a)). 
For $\omega_{c1} \leq \omega \leq \omega_{c2} (u)$ the chain remains in the
form of globule attached to the surface. In this range the monomer-monomer
attraction remains effective in holding the monomers in the neighbourhood
of each other than the surface-monomer attraction whose tendency is to
spread the chain on the surface (see Fig. 2(b)). For $\omega > 
\omega_{c2} (u)$ the globule conformation becomes unstable as surface-monomer
attraction becomes more effective than the monomer-monomer attraction 
and therefore the chain spreads over the surface (just like a liquid 
spreads over a wetting surface).

\vspace {-.2in}

\psfig{figure=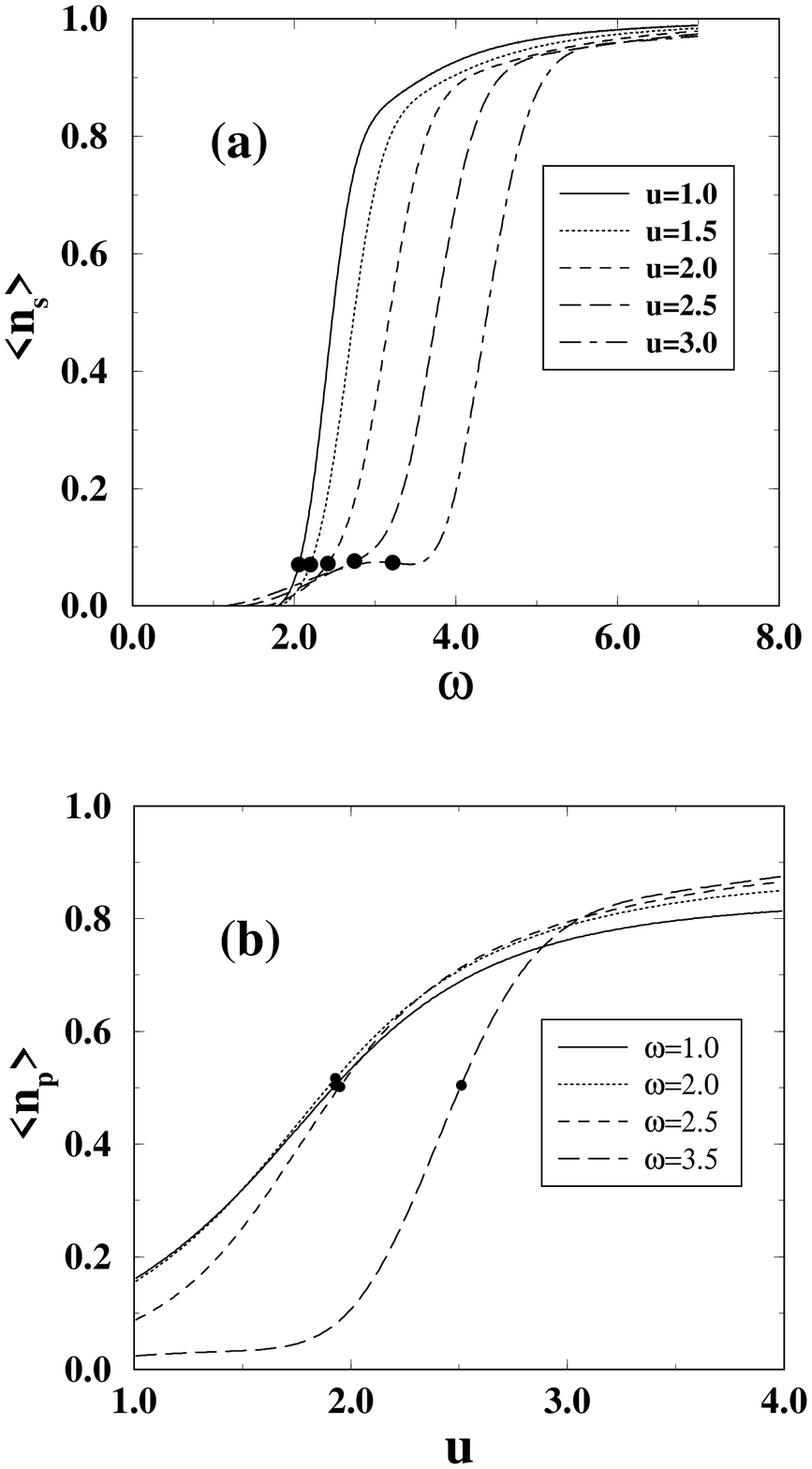,height=5in,width=4in}

\vspace {-.2in}

{\bf FIG.3: (a) \small Average fraction of monomers on the surface ($<n_s>$)
as a function of $\omega$ and  (b) number of pairs ($<n_p>$) as a function
of $u$. Dot on a curve indicates the transition point; in (a) adsorption -
desorption and in (b) expanded - collapsed transitions.}

\vspace {.2in}

Summarizing we studied a SASAW in the presence of an attracting 
impenetrable wall and obtained the phase boundaries separating different 
phases of the polymer chain from data obtained by exact enumerations.
We report a new adsorbed state which has conformation of a compact
globule sticking to a surface in same way as a liquid drop may
lie on a non wetting surface.  The monomer density distribution, 
the number of monomers on the surface and the number of nearest neighbours 
in different regimes of the phase diagram are obtained.

The work was supported by the Department of Science and Technology 
(India) through a project grant.

\end{document}